\documentclass[english]{article}

\usepackage{microtype}
\usepackage{graphicx}
\usepackage{subfigure}
\usepackage{booktabs} 

\usepackage{hyperref}

\usepackage[T1]{fontenc}
\usepackage[latin9]{inputenc}
\usepackage{babel}
\usepackage{array}
\usepackage{float}
\usepackage{multirow}
\usepackage{amsmath}
\usepackage{amssymb}
\usepackage{graphicx}
\usepackage[numbers]{natbib}

\usepackage[accepted]{icml2019}

\icmltitlerunning{Achieving Conservation of Energy in Neural Network Emulators for Climate Modeling}

\begin{document}

\twocolumn[
\icmltitle{Achieving Conservation of Energy in Neural Network Emulators for\\ Climate Modeling}



\icmlsetsymbol{equal}{*}

\begin{icmlauthorlist}
\icmlauthor{Tom Beucler}{equal,ESS,EEE}
\icmlauthor{Stephan Rasp}{LMU}
\icmlauthor{Michael Pritchard}{ESS}
\icmlauthor{Pierre Gentine}{EEE}
\end{icmlauthorlist}

\icmlaffiliation{ESS}{Department of Earth System Science, University of California, Irvine, CA, USA}
\icmlaffiliation{EEE}{Department of Earth and Environmental Engineering, Columbia University, New York, NY, USA}
\icmlaffiliation{LMU}{Meteorological Institute, Ludwig-Maximilian-University, Munich, Germany}

\icmlcorrespondingauthor{Tom Beucler}{tom.beucler@gmail.com}

\icmlkeywords{neural network, climate, energy, conservation, deep learning, climate change, constraints}

\vskip 0.1in
]




\begin{abstract}
Artificial neural-networks have the potential to emulate cloud processes with higher accuracy than the semi-empirical emulators currently used in climate models. However, neural-network models do not intrinsically conserve energy and mass, which is an obstacle to using them for long-term climate predictions. Here, we propose two methods to enforce linear conservation laws in neural-network emulators of physical models: Constraining (1) the loss function or (2) the architecture of the network itself. Applied to the emulation of explicitly-resolved cloud processes in a prototype multi-scale climate model, we show that architecture constraints can enforce conservation laws to satisfactory numerical precision, while all constraints help the neural-network better generalize to conditions outside of its training set, such as global warming. 
\end{abstract}
\section{Motivation}

The largest source of uncertainty in climate projections is the response
of clouds to warming \citep{Schneider2017}. The turbulent eddies generating
clouds are typically only $\mathrm{{\cal O}\left(100m-10km\right)}\ $-wide,
meaning that climate models need to be run at spatial resolutions
as fine as $\mathrm{{\cal O}\left(1\mathrm{km}\right)\ }$to prevent large
biases. Unfortunately, computational resources currently limit climate
models to spatial resolutions of $\mathrm{{\cal O}\left(25\mathrm{km}\right)}\ $when
run for time periods relevant to societal decisions, e.g. 100 years \citep{IPCC2014}. Therefore, climate models rely on semi-empirical
models of cloud processes, referred to as \textit{\small{}convective
parametrizations} \citep{Stevens2013,Sherwood2014}. If designed by
hand, convective parametrizations are unable to capture the complexity
of cloud processes and cause well-known biases, including a lack of
extreme precipitation events and unrealistic cloud structures \citep{Daleu2015a,Daleu2016}.

Recent advances in statistical learning offer the possibility of designing
data-driven convective parametrizations by training algorithms on
short-period but high-resolution climate simulations \citep{Gentine2018a}.
The first attempts have successfully modeled the interaction between
small-scale clouds and the large-scale climate, offering a pathway
to improve the accuracy of climate predictions \citep{Brenowitz2018,Rasp2018,Krasnopolsky2013}.
However, machine learning-based climate models do not intrinsically
conserve energy and mass, which is a major obstacle to their adoption
by the physical science community for several reasons, e.g.:

1) Realistic simulations of climate change respond to relatively small
$\mathrm{{\cal O}\left(1W\ m^{-2}\right)}\ $radiative forcing from carbon
dioxide. Inconsistencies of this magnitude can prevent this
small forcing from being communicated down to the surface and the
ocean where most of the biomass lives.

2) Artificial sources and sinks of mass and energy distort weather
and cloud formation on short timescales, resulting in large temperature
and humidity drifts or biases for the long-term climate.

Current machine-learning convective parametrizations that conserve
energy are based on decision trees (e.g. random forests), but these are too slow for practical use in climate models \citep{Ogorman}. Since neural-network convective parametrizations can significantly reduce cloud biases in climate
models while decreasing their overall computational cost \citep{Rasp2018}, we ask: How can we enforce conservation laws in neural-network emulators of physical models? 

After proposing two methods to enforce physical constraints in neural
network models of physical systems in Section \ref{sec:Theory}, we apply them to emulate cloud processes in a climate model in Section \ref{sec:Application_cloud_climate}, before comparing their performances and how they improve climate predictions
in Section \ref{sec:Results}.

\section{Theory\label{sec:Theory}}

\begin{figure}[ht]
\centerline{\includegraphics[width=0.6\columnwidth]{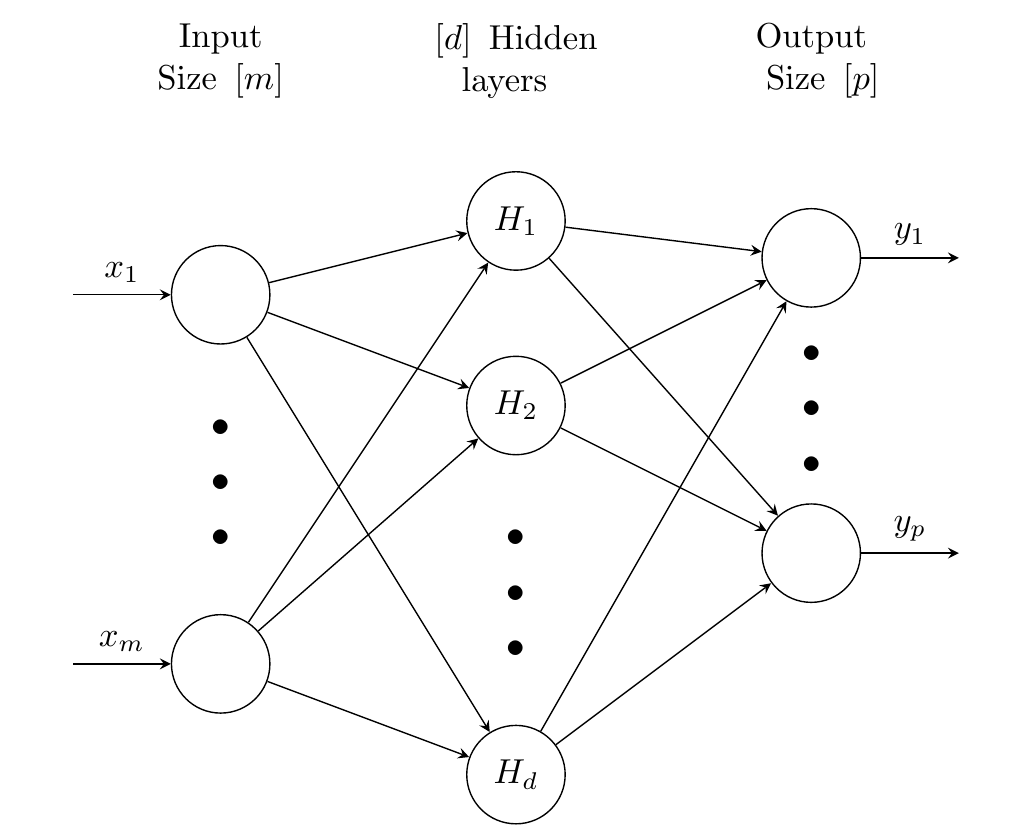}}
\caption{Standard feed-forward configuration $\left(\mathrm{NN}\right)$\label{fig:NN}}
\end{figure}

Consider a physical system represented by a function $f:\ \mathbb{R}^{m}\mapsto\mathbb{R}^{p}\ $that
maps an input $x\in\mathbb{R}^{m}$ to an output $y\in\mathbb{R}^{p}$:
\begin{equation}
y=f\left(x\right).
\end{equation}
Many physical systems satisfy exact physical constraints, such as the
conservation of energy or momentum. In this paper, we assume that
these physical constraints $\left({\cal C}\right)\ $can be written
as an under-determined linear system of rank $n$:
\begin{equation}
\left({\cal C}\right)\overset{\mathrm{def}}{=}\left\{ \boldsymbol{C}\left[\begin{array}{c}
x\\
y
\end{array}\right]=0\right\} ,\label{eq:Conservation}
\end{equation}
where $\boldsymbol{C}\in\mathbb{R}^{n}\times\mathbb{R}^{m+p}\ $is
a constraints matrix acting on the input and output of the system.
The physical system has $n\ $constraints, and by construction, $n<p+m$.
Our goal is to build a computationally-efficient emulator of the physical
system $f\ $and its physical constraints $\left({\cal C}\right)$.
For the sake of simplicity, we build this emulator using a feed-forward
neural network $\left(\mathrm{NN}\right)\ $trained on preexisting
measurements of $x\ $and $y$, as shown in Figure \ref{fig:NN}.
 We measure the quality of $\left(\mathrm{NN}\right)\ $using the
mean-squared error, defined as: 
\begin{equation}
\mathrm{MSE}\left(y,y_{\mathrm{NN}}\right)\overset{\mathrm{def}}{=}\left\Vert y-y_{\mathrm{NN}}\right\Vert \overset{\mathrm{def}}{=}\frac{1}{p}\sum_{i=1}^{p}\left(y_{i}-y_{\mathrm{NN,}i}\right)^{2},\label{eq:MSE}
\end{equation}
where $y_{\mathrm{NN}}\ $is the neural network's output and $y\ $the
``truth''. Our reference case, referred to as ``unconstrained
neural network'' (NNU), optimizes $\left(\mathrm{NN}\right)\ $using
$\mathrm{MSE}\ $as its loss function. To enforce the physical constraints
$\left({\cal C}\right)\ $in our neural network, we consider two options:
\begin{enumerate}
\item \textbf{Constraining the loss function $\left(\mathrm{NNL}\right)$}:
In this setting, we penalize our neural network for violating physical
constraints using a penalty ${\cal P}$, defined as the residual from
the physical constraints:

\begin{equation}
{\cal P}\left(x,y_{\mathrm{NN}}\right)\overset{\mathrm{def}}{=}\left\Vert \boldsymbol{C}\left[\begin{array}{c}
x\\
y_{\mathrm{NN}}
\end{array}\right]\right\Vert.
\label{eq:Penalty}
\end{equation}

We apply this penalty by giving it a weight $\alpha\in[0,1]\ $in
the loss function ${\cal L}$, which is similar to a Lagrange multiplier: 
\begin{equation}
{\cal L}\left(\alpha\right)=\alpha{\cal P}\left(x,y_{\mathrm{NN}}\right)+\left(1-\alpha\right)\mathrm{MSE}\left(y,y_{\mathrm{NN}}\right).
\end{equation}

\item \textbf{Constraining the architecture $\left(\mathrm{NNA}\right)$:}
In this setting, we augment the simple network $\left(\mathrm{NN}\right)\ $with
$n\ $conservation layers to enforce the conservation laws $\left({\cal C}\right)\ $to
numerical precision (Figure \ref{fig:NNA}), while still calculating the $\mathrm{MSE}\ $loss over the \textit{entire} output vector $y$. The feed-forward network outputs an ``unconstrained''
vector $u\in\mathbb{R}^{p-n}$ whose size is only $\left(p-n\right)$,
where $n\ $is the number of constraints. We then calculate the remaining
component $v\in\mathbb{R}^{n}\ $of the output vector $y_{\mathrm{NN}}\ $using
the $n\ $constraints. This defines $n\ $constraint layers $\left(\mathrm{CL}_{1..n}\right)\ $that
ensure that the final output $y_{\mathrm{NN}}\ $exactly respects
the physical constraints $\left({\cal C}\right)$. A possible construction
of $\left(\mathrm{CL}_{1..n}\right)\ $solves the system of equations
$\left({\cal C}\right)\ $from the bottom to the top row after writing
it in row-echelon form. Note that the loss is propagated through the physical constraints.

\begin{figure}[ht]
\centerline{\includegraphics[width=0.7\columnwidth]{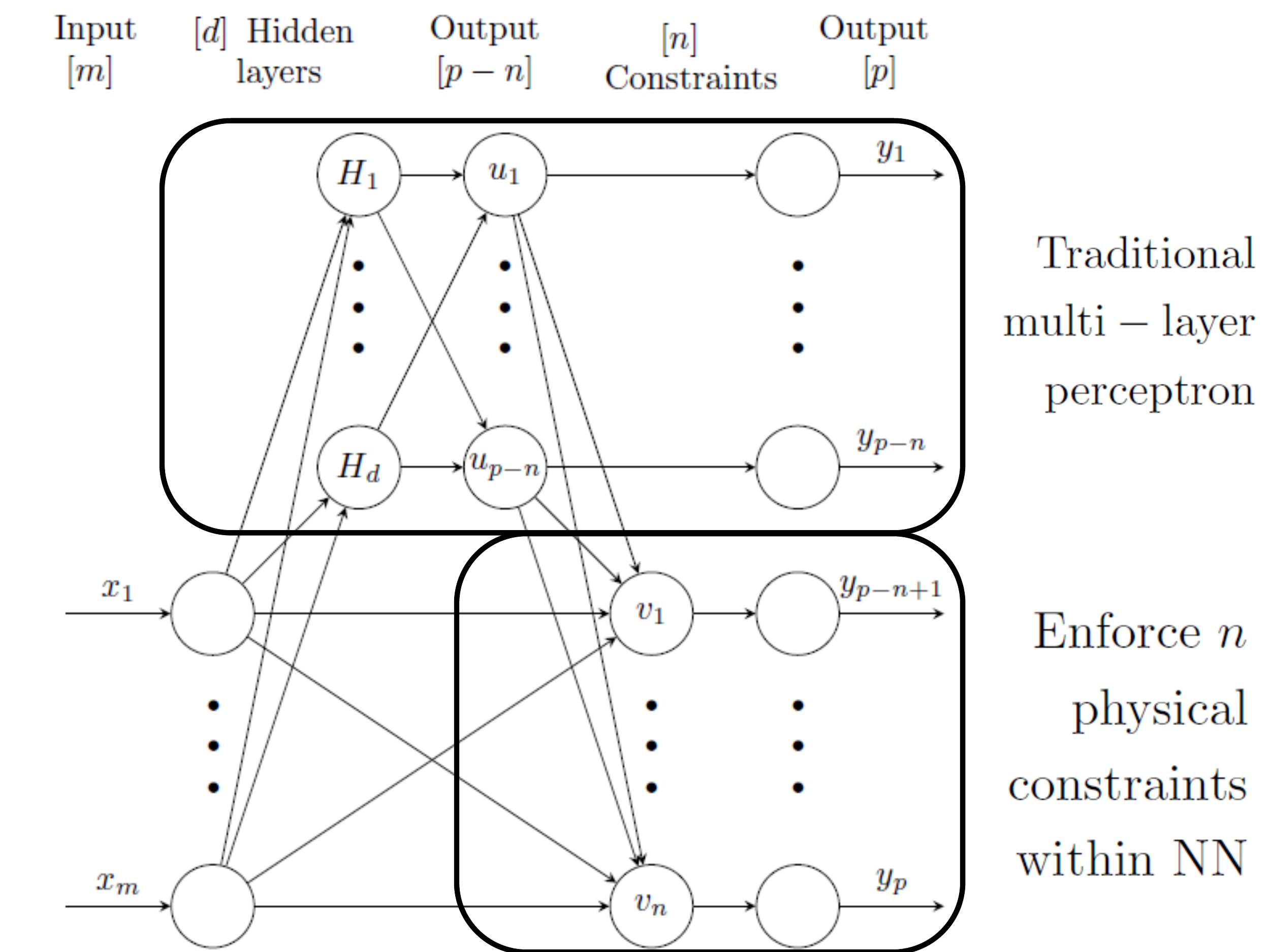}}
\caption{Architecture-constrained configuration $\left(\mathrm{NNA}\right)$\label{fig:NNA}}
\end{figure}
 
\end{enumerate}

\section{Application to Convective Parametrization for Climate Modeling\label{sec:Application_cloud_climate}}

\begin{table*}[t]
\begin{centering}
{\small{}}%
\begin{tabular}{c|c|ccccc}
{\small{}Validation} & {\small{}Metric} & {\small{}Linear (MLR)} & {\small{}Uncons. (NNU)} & {\small{}Loss ($\mathrm{NNL}_{\alpha=0.01}$)} & {\small{}Loss ($\mathrm{NNL}_{\alpha=0.5}$)} & {\small{}Architecture (NNA)}\tabularnewline
\hline 
{\small{}Baseline} & {\small{}$\mathrm{MSE}$} & {\small{}$295\pm1.7.10^{3}$} & {\small{}$156\pm1.0\times10^{3}$} & {\small{}$154\pm1.0\times10^{3}$} & {\small{}$177\pm1.1\times10^{3}$} & {\small{}$169\pm1.0\times10^{3}$}\tabularnewline
(+0K) & {\small{}${\cal P}$} & {\small{}$28\pm2\times10^{1}$} & {\small{}$458\pm5\times10^{2}$} & {\small{}$125\pm2\times10^{2}$} & {\small{}$5.0\pm5$} & {\small{}$7\times10^{-10}\pm1\times10^{-9}$}\tabularnewline
\cline{1-1} 
Cl.change & {\small{}$\mathrm{MSE}$} & {\small{}$747\pm1\times10^{5}$} & {\small{}$633\pm7\times10^{3}$} & {\small{}$471\pm5\times10^{3}$} & {\small{}$496\pm8\times10^{3}$} & {\small{}$567\pm8\times10^{3}$}\tabularnewline
(+4K) & {\small{}${\cal P}$} & {\small{}$265\pm2\times10^{3}$} & {\small{}$3\times10^{5}\pm1\times10^{6}$} & {\small{}$2\times10^{3}\pm1\times10^{4}$} & {\small{}$470\pm2\times10^{3}$} & {\small{}$2\times10^{-9}\pm5\times10^{-9}$}\tabularnewline
\hline 
\end{tabular}
\par\end{centering}{\small \par}

\caption{Mean-Squared Error (skill) and Physical Constraints Penalty ${\cal P}$ (violation of energy/mass/radiation conservation laws) for
different neural networks in units $\mathrm{W^{2}\ m^{-4}}\ $using
the format $\left(\mathrm{Mean}\pm\mathrm{Standard\ deviation}\right)$.\label{tab:MSE_P}}
\end{table*}

\begin{figure*}[t]
\centerline{\includegraphics[width=14cm]{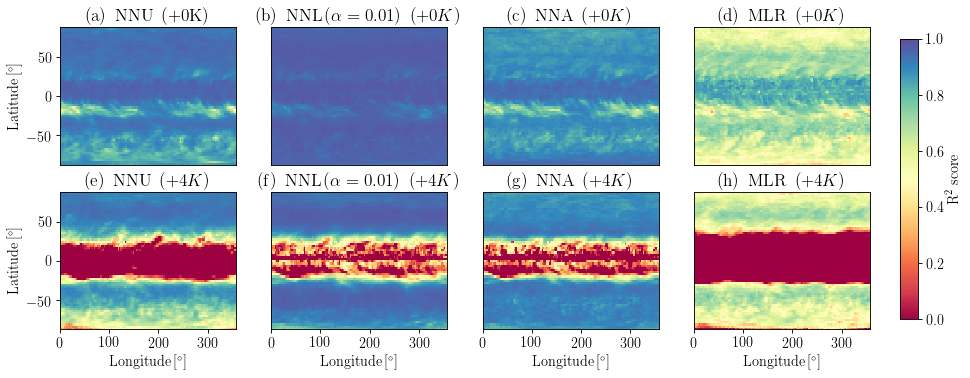}}

\caption{$R^{2}\ $scores of different neural networks simulating the outgoing
longwave radiation field over the entire planet for the (+0K) dataset
(first row) and (+4K) dataset (second row).\label{fig:R2_OLR}}
\end{figure*}



We now implement the three neural networks $\left(\mathrm{NNU,NNL,NNA}\right)\ $and
compare their performances in the particular case of convective parametrization via emulation of the 8,192 cloud-resolving sub-domains embedded in the Super-Parametrized Community Atmosphere Model 3.0 \citep{Collins2006,Khairoutdinov2005}.
We simulate an ``ocean world'' where the surface temperatures are fixed with
a realistic equator-to-pole gradient \citep{Andersen2012}. To facilitate
the comparison, all networks have 5 hidden layers with 512 nodes each,
and use leaky rectangular unit activation functions: $x\mapsto\max\left(0.3x,x\right)\ $to
help capture the system's non-linearity. We use the RMSprop optimizer
\citep{tieleman2012lecture} to train each network during 20 epochs,
using 3 months of climate simulation with 30-minute outputs as training
data.

The goal of the neural network is to predict an output vector $y\ $of
size 218 that represents the effect of cloud processes on climate (i.e. convective and radiative tendencies),
 based on an input vector $x\ $of size 304 that represents the climate
state (i.e. large-scale thermodynamic variables).
The 4 conservation laws can be written as a sparse matrix of size
$4\times\left(304+218\right)\ $that acts on $x\ $and $y\ $to yield
equation \ref{sec:Theory}.

Each row of the conservation matrix $\boldsymbol{C}\ $describes a
different conservation law: The first row is the conservation of enthalpy,
the second row is the conservation of mass, the third row is the conservation
of terrestrial radiation and the last row is the conservation of solar
radiation. In the architecture-constrained case, we output an unconstrained
vector $u\ $of size $\left(218-4\right)=214$, and calculate the
4 remaining components $v\ $of the output vector $y\ $by solving
the system of equations $\boldsymbol{C}\left[\begin{array}{cc}
x & y\end{array}\right]^{T}=0\ $from bottom to top. 


We evaluate the performances of $\left(\mathrm{NNU,NNL,NNA}\right)$
on two different validation datasets:

(+0K) An ``ocean world'' similar to the training dataset.

(+4K) An ``ocean world'' where the surface temperature has been uniformly warmed by 4K, a proxy for the effects of climate
change. We do not expect (NN) to perform well in the Tropics, where this perturbation leads to temperatures outside of the training set.

\section{Results\label{sec:Results}}

Table \ref{tab:MSE_P} compares the performance and the degree to
which each neural network violates conservation laws, as measured
by the mean-squared error and the penalty ${\cal P}$, respectively.

 All neural networks perform better than the multiple-linear regression
model (MLR), derived by replacing leaky rectangular units with the
identity function and optimized independently. While the reference ``unconstrained''
network NNU performs well as measured by $\mathrm{MSE}$, it does
so by breaking conservation laws, resulting in a large penalty ${\cal P}$.
Enforcing conservation laws via architecture constraints (NNA) works
to satisfactory numerical precision on both validation datasets, resulting in a
very small penalty ${\cal P}$. Giving equal weight to $\mathrm{MSE}\ $and
${\cal P}\ $in the loss function$\ \left(\mathrm{NNL}_{\alpha=0.5}\right)\ $leads
to mediocre performances in all areas. In contrast, surprisingly, introducing the
penalty ${\cal P}\ $in the loss function with a very small weight
$\left(\alpha=0.01\right)$ leads to the best performance on the reference
validation dataset $\left(+0K\right)\ $. Both constrained networks $\mathrm{NNL}_{\alpha=0.01}\ $ and $\mathrm{NNA}$ generalize better to unforeseen conditions $\left(+4K\right)$ than the "unconstrained" network, suggesting that physically constraining neural networks improves their representation abilities. This ability to generalize is confirmed by the high $R^{2}-$score when predicting the outgoing
longwave radiation (Figure \ref{fig:R2_OLR}), which can be used as
a direct measure of radiative forcing in climate change scenarios.

Overall, our results suggest that (1) constraining the network's
architecture is a powerful way to ensure energy conservation over
a wide range of climates and (2) introducing a very small information
about physical constraints in the loss function or/and the network's architecture can significantly
improve the generalization abilities of our neural
network emulators.





\bibliography{2019_ICML_BIB.bib}
\bibliographystyle{plainnat}





\end{document}